\documentclass[10pt, twocolumn, pre, aps, superscriptaddress, showpacs]{revtex4-1}
\usepackage{amsmath, graphicx, subfigure, tikz}
\begin{document}

\title{Complete Density Calculations of q-State Potts and Clock Models:\\
Reentrance of Interface Densities under Symmetry Breaking}
    \author{E. Can Artun}
    \affiliation{Faculty of Engineering and Natural Sciences, Kadir Has University, Cibali, Istanbul 34083, Turkey}
    \author{A. Nihat Berker}
    \affiliation{Faculty of Engineering and Natural Sciences, Kadir Has University, Cibali, Istanbul 34083, Turkey}
    \affiliation{Department of Physics, Massachusetts Institute of Technology, Cambridge, Massachusetts 02139, USA}



\begin{abstract}

All local bond-state densities are calculated for $q$-state Potts
and clock models in three spatial dimensions, $d=3$.  The
calculations are done by an exact renormalization group on a
hierarchical lattice, including the density recursion relations, and
simultaneously are the Migdal-Kadanoff approximation for the cubic
lattice. Reentrant behavior is found in the interface densities
under symmetry breaking, in the sense that upon lowering temperature
the value of the density first increases, then decreases to its zero
value at zero temperature.  For this behavior, a physical mechanism
is proposed. A contrast between the phase transition of the two
models is found, and explained by alignment and entropy, as the
number of states $q$ goes to infinity.  For the clock models, the
renormalization-group flows of up to twenty energies are used.

\end{abstract}
\maketitle

\section{Introduction: Total Renormalization-Group Solution of Two Families of Models}

Although originally introduced for critical phenomena,
renormalization-group calculation gives the total thermodynamics of
a system, at and away phase transitions \cite{BerkerOstlund}.  In
order to effect this, the recursion relations of the local densities
are needed, leading to calculation more complicated than that for
phase boundaries and critical exponents.  This calculation is
carried out here for two families of models, namely Potts
\cite{BerkerPLG, NienhuisPotts, AndelmanBerker} and clock
\cite{BerkerNelson}, each with $q$ states, on a hierarchical lattice
\cite{BerkerOstlund,Kaufman1, Kaufman2} in three spatial dimensions,
$d=3$. The calculation is exact for the hierarchical and is
considered approximate for a cubic lattice \cite{Migdal,Kadanoff}.
The temperature functions and symmetry-breaking behaviors of dozens
local densities are derived and interesting behaviors are found, and
explained, such as a reentrance behavior in the interface densities.
The models are similarly defined, but exhibit different behaviors,
such as the $q$ saturation of the magnetization and the phase
transitions as $q$ goes to infinity, which is also explained.

\section{Potts and Clock, and Densities Calculation}

\subsection{The $q$-State Models and Their Set of Densities}

These general $q$-state models are simply defined by the
Hamiltonians, for the Potts models,
\begin{equation}
- \beta {\cal H} = \sum_{\left<ij\right>} \, J\delta(s_is_j),
\end{equation}
where $\beta=1/k_{B}T$, at site $i$ the spin $s_{i}=a,b,,...,$ can
be in $q$ different states, the delta function $\delta(s_i
s_j)=1(0)$ for $s_i=s_j (s_i\neq s_j)$, and $\langle ij \rangle$
denotes summation over all nearest-neighbor pairs of sites.  For the
clock models,
\begin{equation}
- \beta {\cal H} = \sum_{\left<ij\right>} \, J
\cos(\overrightarrow{s}_i \cdot \overrightarrow{s}_j),
\end{equation}
where at site $i$ the spin $\overrightarrow{s}_{i}$ can point in $q$
different directions $\theta _i = 2\pi n_i/q$ in the $xy$ plane,
with $n_{i}=0,1,...,q-1$ providing the $q$ different possible
states. The limit $q \rightarrow \infty$ of the clock model gives
the $XY$ model, which we also explore here, with results (physically
explainably) quite different from the $q \rightarrow \infty$ limit
of the Potts model (Fig. 1).
\begin{figure}[ht!]
\centering
\includegraphics[scale=0.23]{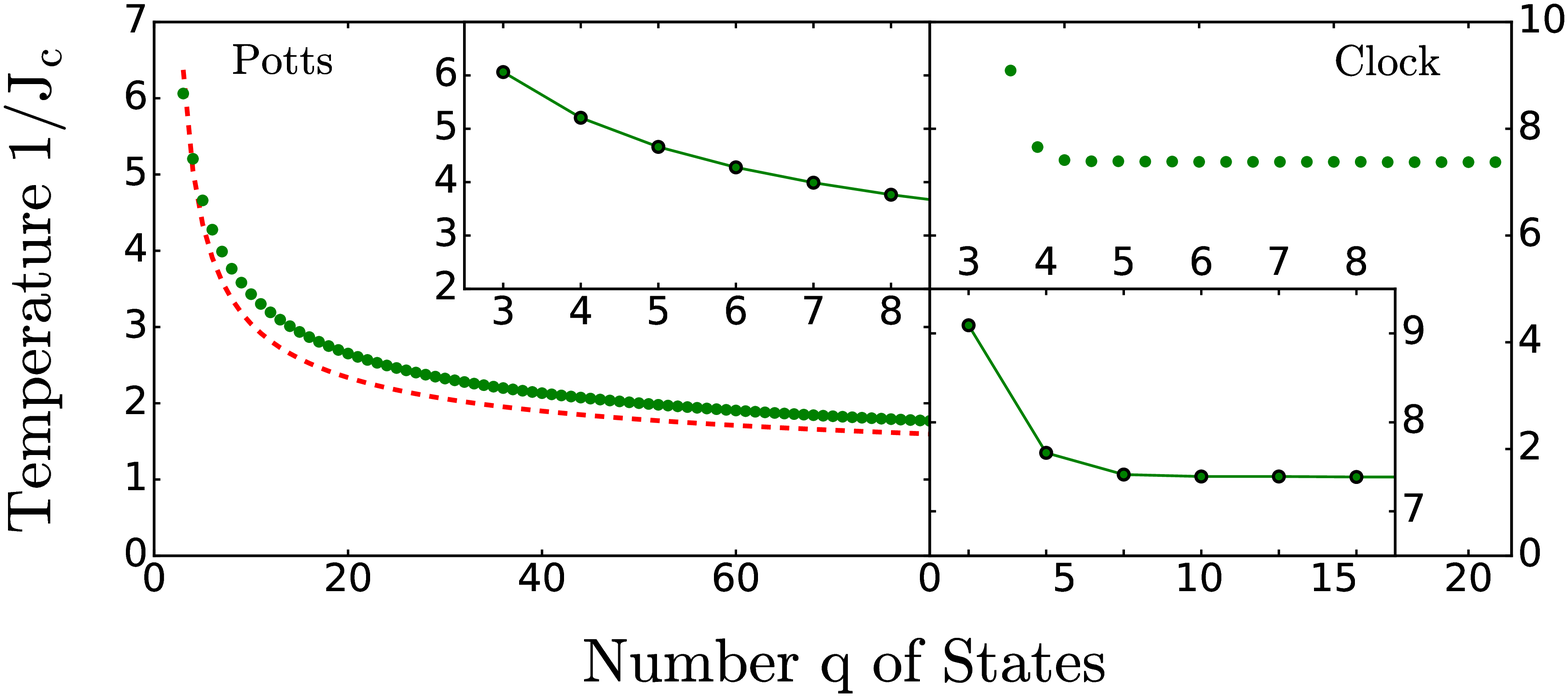}
\caption{Calculated critical temperatures $J^{-1}_c$ of Potts and
clock models as a function of number of states $q$, in $d=3$.  From
this figure and from Table I, it is seen that the clock model
quickly (at as low as $q=5$) settles to its $q=\infty$ (which is the
$XY$ model) value of $J^{-1}_c=7.4$ . The dashed line for the Potts
critical temperatures is $J^{-1}_c = 7/\ln(q),$ derived here for
strong coupling.}
\end{figure}

Our aim is to calculate all (there are $q(q+1)/2$ of them) of the
bond-state densities
\begin{equation}
U(n_in_j) = <\delta(s_in_i) \delta(s_jn_j)>,
\end{equation}
where $(i,j)$ are the sites on each end of the bond and $n_i$
designates one of $q$ possible states of the spin $s_i$.  These
bond-state densities are obtained from the partition function $Z$,
\begin{equation}
U(n_in_j) = \frac{1}{N} \frac{\partial \ln Z}{\partial E(n_in_j)},
\end{equation}
where $N$ is the number of nearest-neighbor pairs in the system and
$E(n_in_j)$ is the energy assigned to the bond when its sites are in
states $(n_i,n_j)$.  Before any renormalization, these bond energies
are given by Eqs. (1)and (2),
\begin{equation}
E(n_in_j) = J\delta(n_in_j) \quad \text{and} \quad J \cos(2\pi
(n_i-n_j)/q),
\end{equation}
for Potts and clock models, respectively. The $E(n_in_j)$ are the
(large number of, see below) renormalization-group flow variables
and Eqs. (5) give the initial conditions, parametrized by
temperature $J^{-1}$, of the renormalization-group flows.  The forms
in Eqs. (5) are of course not conserved during the flows.

\subsection{Energy Recursion Relations of the Renormalization Group}

For our renormalization-group calculation, we use the
Migdal-Kadanoff approximation, which, as shown in Fig. 2(a),
consists in bond-moving followed by decimation
\cite{Migdal,Kadanoff}.  This operation is equivalent to
constructing the $q \times q$ transfer matrix $T(n_in_j)=
\exp(E(n_in_j))$, taking the $b^{d-1}$ th power of each element of
the matrix (this is bond moving) and matrix multiplying the
resulting matrix with itself (this is decimation). For numerical
convenience at the low-temperature sink of the flows, after every
decimation (and before first starting the first renormalization), we
subtract $E(aa)$ (for Potts) or $E(00)$ (for clock) from all
$E(n_in_j)$, thus setting $E(aa)=0$ or $E(00)=0$ and introducing the
additive constant $NG$ in the Hamiltonian, which has the
renormalization-group recursion relation
\begin{equation}
G' = b^d G + \tilde{G},
\end{equation}
where, here and everywhere, prime refers to the renormalized system,
the first term is the additive constants the renormalized bond
inherits from the $b^d$ bonds it replaces and the second term comes
from compensating the subtraction of $E(aa)$ or $E(00)$. These
recursion relations are then in terms of the elements (or
equivalently their logarithms $E(n_in_j) = \ln (T(n_in_j)$)  of the
diagonal and upper left triangle of the transfer matrix (since this
matrix is symmetrical). The number of these elements can be somewhat
reduced by noting those identically equal by symmetry and not to be
distinguished by possible spontaneous symmetry breaking, as is
illustrated for the clock models below, but it will be seen that the
number of the flow variables for the clock models rapidly increases
with $q$.  A large $q$ calculation, such as the one we do here for
$q=360$ to probe the $q \rightarrow \infty$ $XY$ model limit, is
best effected by doing directly numerically the matrix operations
described above on the $360 \times 360$ transfer matrix.  By
contrast, for any $q$, by using the (partially broken under
ordering) permutation symmetry of the Potts variables, we can reduce
the number of renormalization-group flow variables to 4, which makes
it possible to treat any $q$, including $q=\infty$, as seen below.
The recursion relations obtained by the Migdal-Kadanoff
approximation are exactly applicable to the exact solution of the
hierarchical lattice shown in Fig. 2(b)
\cite{BerkerOstlund,Kaufman1,Kaufman2}. Thus, a ''physically
realizable'', therefore robust approximation is used. Physically
realizable approximations have been used in polymers
\cite{Flory,Kaufman}, disordered alloys \cite{Lloyd}, and turbulence
\cite{Kraichnan}.  Recent works using exactly soluble hierarchical
lattices are in Refs.
\cite{Monthus,Sariyer,Ruiz,Rocha-Neto,Ma,Boettcher5}.

\begin{figure}[ht!]
\centering
\includegraphics[scale=0.4]{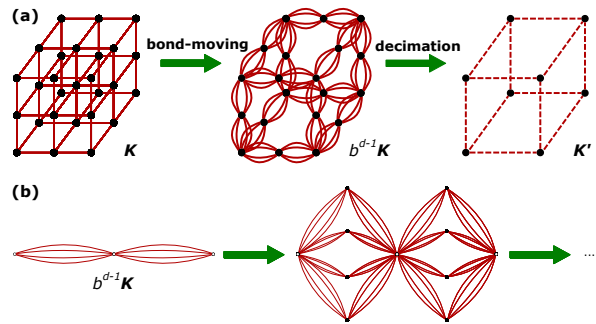}
\caption{(a) Migdal-Kadanoff approximate renormalization-group
transformation for the $d=3$ cubic lattice with the length-rescaling
factor of $b=2$. (b) Construction of the $d=3, b=2$ hierarchical
lattice for which the Migdal-Kadanoff recursion relations are exact.
The renormalization-group solution of a hierarchical lattice
proceeds in the opposite direction of its construction.}
\end{figure}
\begin{figure}[ht!]
\centering
\includegraphics[scale=0.28]{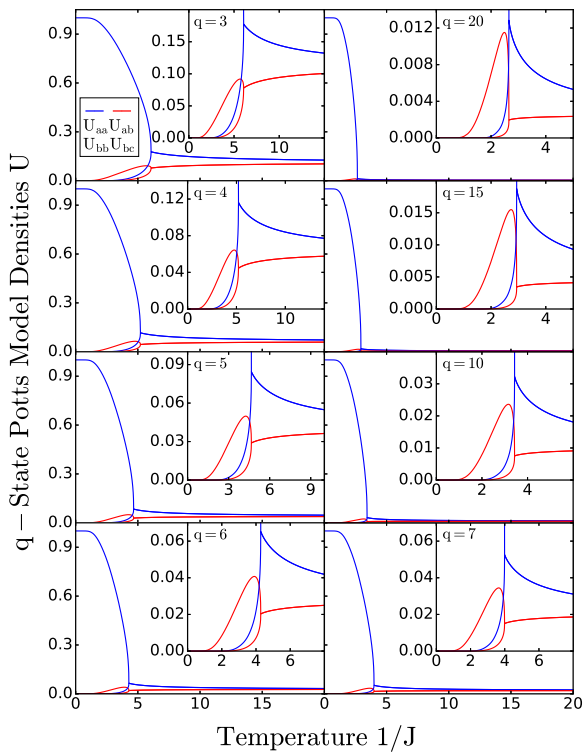}
\caption{The calculated nearest-neighbor densities of the $q$-state
Potts models in $d=3$.  The upper curve on the right is $U_{aa}$ and $U_{bb}$,
which coincide in the disordered high-temperature phase and split in
the low-temperature phase where symmetry is spontaneously broken in
favor of state $a$. The lower curve on the right is $U_{ab}$ and $U_{bc}$, also
which coincide in the disordered high-temperature phase and split in
the symmetry-broken low-temperature phase. The interface density
$U_{ab}$ exhibits reentrance as temperature is lowered in the
ordered phase, first increasing in value and then receding to zero
at zero temperature.}
\end{figure}
\begin{figure}[ht!]
\centering
\includegraphics[scale=0.23]{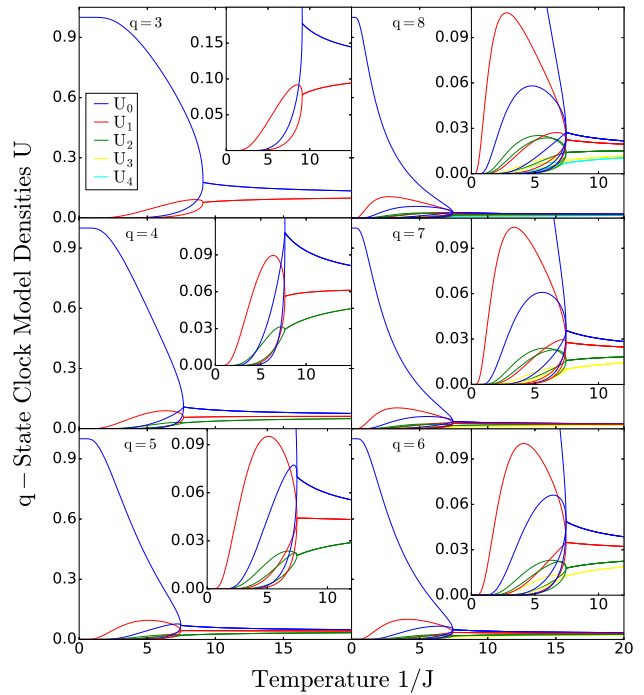}
\caption{The calculated nearest-neighbor densities of the $q$-state
clock models in $d=3$.  The curves are for $U_{m} \equiv U_{k,k-m}$,
for $k=0,1,...,q-1$ and $m=0,1,...,$ from the top to downwards in
each figure panel. Thus, $m$ measures the angular difference $\theta
_i - \theta _j = 2\pi m/q$ between the states of neighboring spins.
For each $m$, the curves for different $k$ coincide in the
disordered high-temperature phase. In the low-temperature phase, for
each $m$, the densities involving $k=0$ and the densities involving
$k>0$ split under the symmetry breaking favoring the state 0. The
interface densities involving $k=0$ exhibit reentrance as
temperature is lowered in the ordered phase, first increasing in
value and then receding to zero at zero temperature.}
\end{figure}
\begin{figure}[ht!]
\centering
\includegraphics[scale=0.22]{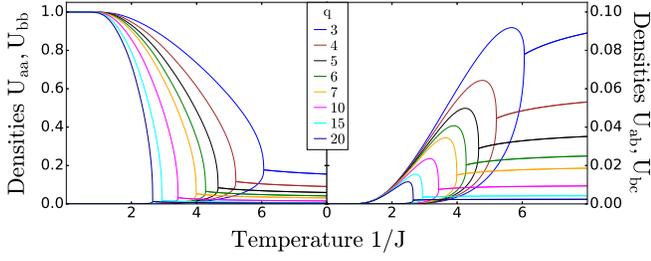}
\caption{Comparison with respect to the number of states
$q=3,4,5,6,7,10,15,20$ from top down in each panel, of the
nearest-neighbor densities of the Potts models in $d=3$.  The right
panel shows the curves for neighboring unlike states $U_{ab}$ and
$U_{bc}$, which coincide in the disordered high-temperature phase
and split in the symmetry-broken low-temperature phase. The
interface density $U_{ab}$ exhibits reentrance as temperature is
lowered in the ordered phase, first increasing in value and then
receding to zero at zero temperature. This reentrance is pronounced
in the low $q$ states and decreases for high $q$. The left panel
shows the curves for the like-state neighbors $U_{aa}$ and $U_{bb}$,
which also coincide in the disordered high-temperature phase and
split in the low-temperature phase where symmetry is spontaneously
broken in favor of state $a$. }
\end{figure}
\begin{figure}[ht!]
\centering
\includegraphics[scale=0.21]{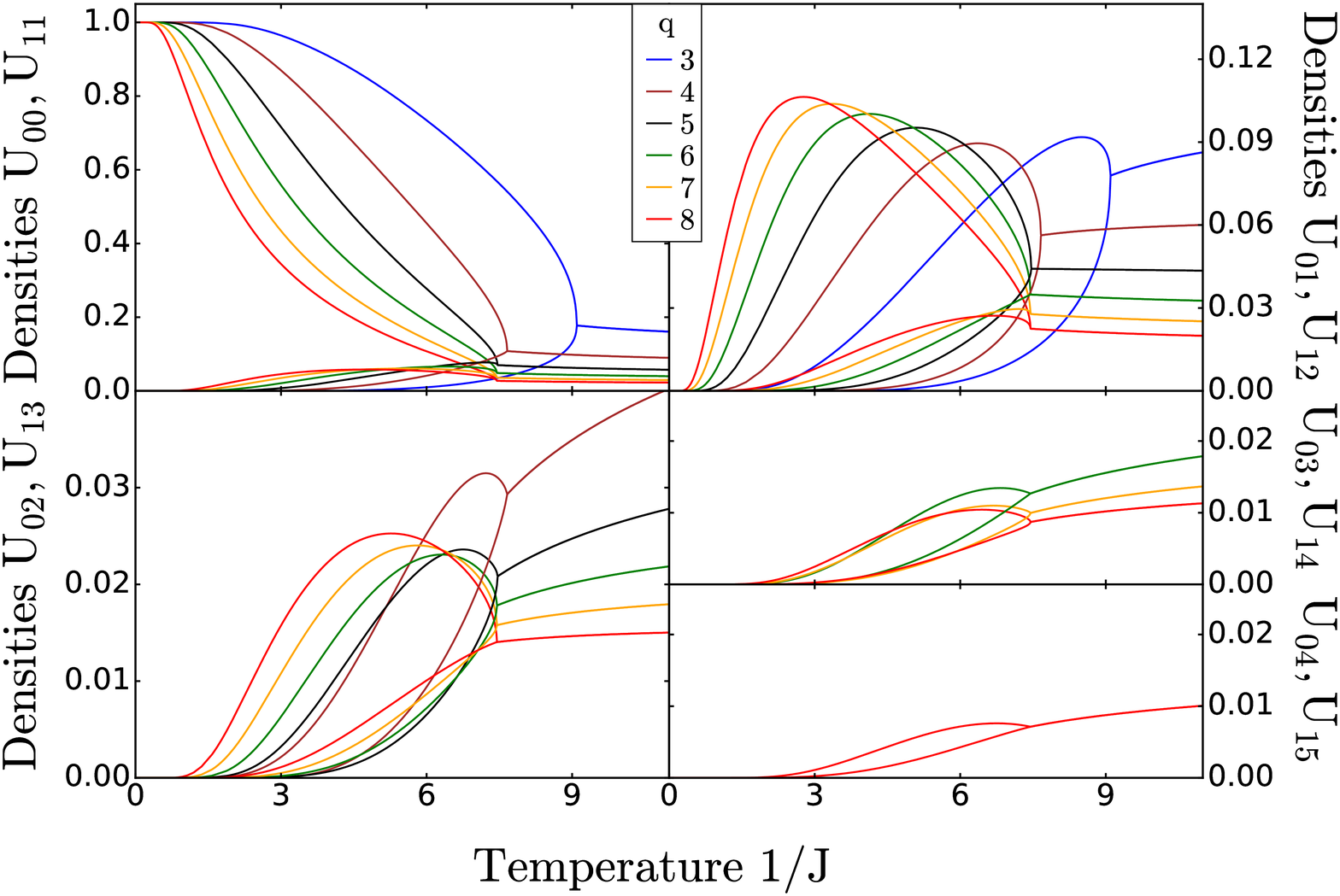}
\caption{Comparison with respect to the number of states
$q=3,4,5,6,7,8$ of the nearest-neighbor densities of the clock
models, $U_{m} \equiv U_{k,k-m}$ for $k=0,1$ and $m=0,1,...,$ shown
in decreasing $q$ on the high-temperature side in each panel.  $m$
measures the angular difference $\theta _i - \theta _j = 2\pi m/q$
between neighboring spins.  The top right panel shows the curves for
neighboring unlike states with $m=1$.  The bottom left panel shows
the curves with $m=2$, and therefore $q=4,5,6,7,8$.  The bottom
right upper panel shows the curves with $m=3$, and therefore with
$q=6,7,8$. The bottom right lower panel shows the curves with $m=4$,
and therefore $q=8$. For each $m$, the curves for different $k$
coincide in the disordered high-temperature phase. In the
low-temperature phase, for each $m$, the densities involving $k=0$
and the densities involving $k>0$ split under the symmetry breaking
favoring the state 0. All interface densities involving $k=0$
exhibit reentrance as temperature is lowered in the ordered phase,
first increasing in value and then receding to zero at zero
temperature.}
\end{figure}

\subsection{Density Recursion Relations of the Renormalization Group}

In each renormalization-group transformation, the densities obey the
recursion relation
\begin{equation}
\textbf{U} = b^{-d} \textbf{U'} \cdot \textbf{R},
\end{equation}
where the densities $\textbf{U} \equiv [1,U(n_in_j)]$ are conjugate
to the fields $\textbf{E} \equiv [G,E(n_in_j)]$ and the recursion
matrix is $\textbf{R} = \partial \textbf{E'} /\partial\textbf{E}$. The exact Eq.~(7) is obtained by using the derivative chain rule on $\textbf{U} = (1/N) \partial \ln Z /
\partial\mathbf{E}$, where $Z$ is the partition function and $N$ is the number of nearest-neighbor pairs of spins, and is used to calculate densities from renormalization-group theory \cite{BerkerOstlund,Ilker2,Atalay}. In these defined vectors, the $E(aa)$ or $E(00)$ and $U(aa)$ or
$U(00)$ are missing, since these energies are set to zero by the
additive constant and therefore do not recur.  $U(aa)$ and $U(00)$
are found from the sum rule $\Sigma_{n_i,n_j} U(n_in_j) = 1$.  The
other densities are calculated by iterating Eq.~(7) until a stable
fixed point (sink of the thermodynamic phase) is reached.  The
densities $\textbf{U*}$ at the sink are the left eigenvectors of
$\textbf{R}$ with eigenvalue $b^d$ and conclude the calculation by
insertion to the right-hand side of Eq.~(7). These will be discussed
below specifically for each model. The unstable fixed point dividing
the renormalization-group flows to the phase sinks, parametrized by
$J$, yields the phase transition temperatures given in Fig. 1 and
Table I.

\section{Results: $q$-State Potts Models}
\subsection{Potts Recursion Relations}
Because of the permutation symmetry of the model, namely that given
the $\delta$ function, with respect to a given state, all other
states are equivalent (unlike the clock model involving the product
of slightly or more aligned vectors) the $q \times q$ transfer
matrix manipulations of the recursion relations given above can be
reduced to four simple equations,
\begin{multline}
\begin{split}
e^{E'(a\overline{b})+\widetilde{G}}&= x(a\overline{b}) + x(a\overline{b})x(\overline{b}\overline{b}) +(q-2)x(a\overline{b})x(\overline{bc}),\\
e^{E'(\overline{bb})+\widetilde{G}}&= x(a\overline{b})^2 + x(\overline{bb})^2  +(q-2)x(\overline{bc})^2,\\
e^{E'(\overline{bc})+\widetilde{G}}&= x(a\overline{b})^2 + 2x(\overline{bb})x(\overline{bc}) +(q-3)x(\overline{bc})^2,\\
e^{\widetilde{G}} &= 1 + (q-1)x(a\overline{b})^2,\\
\end{split}
\end{multline}
where the Potts state $a$ has been singled out for possible
spontaneous symmetry breaking, $\overline{b}$ represents
any Potts state which is not $a$, and $\overline{c}$ represents any
Potts state which is not $a$ or the state in $\overline{b}$, and
$x(a\overline{b}) \equiv e^{b^{d-1}E(a\overline{b})}$, etc. In the
latter equation, the factor $b^{d-1}$ represents bond moving and
Eqs. (8) effect the decimation with the bond-moved energies. The
recursion matrix $\textbf{R}$ is the $4 \times 4$ derivative matrix
of Eqs. (8), and the density calculations can be done for any number
of states $q$, including infinity.  By derivative matrix, we mean the derivatives of the renormalized quantities with respect to the unrenormalized quantities.

The recursion relations of Eqs. (8) flow to one of two phase sinks.
On the high temperature side, the sink of the disordered phase is
\begin{equation}
E(aa)^* = E(a\overline{b})^* = E(\overline{bb})^* =
E(\overline{bc})^* = 0,
\end{equation}
where * denotes the fixed point value.  The left eigenvector, with
eigenvalue $b^d$, of the recursion matrix $\textbf{R}$ at this sink
is
\begin{multline}
\textbf{U*} = [1,U(a\overline{b})^*,U(\overline{bb})^*,U(\overline{bc})^*]=\\
[1,<\delta (s_ia) \delta (s_j\overline{b})>+<\delta (s_i\overline{b})\delta (s_ja)>,\\
<\delta (s_i\overline{b})\delta (s_j\overline{b}>, <\delta(s_i\overline{b})\delta(s_j\overline{c})>]=\\
[1,2(q-1)/q^2,(q-1)/q^2,(q-1)(q-2)/q^2].
\end{multline}
Capping with Eq.~(10) from left the repeated applications of Eq.~(7), the densities
$U(a\overline{b}),U(\overline{bb}),U(\overline{bc})$ are obtained
over the entire temperature range of the high-temperature disordered
phase.  Finally,
\begin{equation}
U(ab) = U(a\overline{b})/U^*(a\overline{b}),
\end{equation}
etc. gives the density for a specific pair of states $(a,b)$.

On the low-temperature side, the sink of the ordered phase is
\begin{equation}
E(aa)^* = E(\overline{bb})^* = 0, \quad E(a\overline{b})^* =
E(\overline{bc})^* \rightarrow -\infty.
\end{equation}
A left eigenvector, with eigenvalue $b^d$, of the recursion matrix
$\textbf{R}$ at this sink is
\begin{equation}
\textbf{U*} = [1,0,0,0].
\end{equation}
Calculation, as described after Eq.~(10) above, gives the densities
over the entire temperature range of the low-temperature ordered
phase, showing spontaneous symmetry-breaking in favor of
state $a$.  This result is described in detail in the next
subsection.

Another left eigenvector, with eigenvalue $b^d$, of the recursion
matrix $\textbf{R}$ at this sink is $[1,1,0,0]$.  This eigenvector
gives symmetry breaking in favor of one of the states
$\overline{b}$, namely one of the states which is not $a$.  This
leads so results identical, with the permutation mapping of the
Potts model, to the results involving symmetry breaking in favor of
$a$.  A linear combination of these two degenerate eigenvectors is
of course also an eigenvector with the eigenvalue $b^d$, physically
corresponding to the macroscopic coexistence of differently
symmetry-broken phases.

It noteworthy that throughout the renormalization-group flows,
\begin{equation}
E(aa) = E(\overline{bb}), \qquad E(a\overline{b}) =
E(\overline{bc}).
\end{equation}
However, these interactions have to be distinguished in the
recursion relations, enabling construction of the $4 \times 4$
recursion matrix $\textbf{R}$, to calculate distinctly $U(aa),
U(\overline{bb})$, and see the symmetry breaking.  This calculation
is also going to lead to the full determination of the
magnetization, as seen below.
\subsection{Potts Densities and Interface Density Reentrance}
The calculated nearest-neighbor densities of the $q$-state Potts
models in $d=3$ are given, for $q=3,4,5,6,7,10,15,20$, in Fig. 3.
(For easy comparison, the densities for the clock models are given
in the adjoining Fig. 4.) The upper curve is $U(aa)$ and $U(bb)$,
which coincide in the disordered high-temperature phase and split in
the low-temperature phase where symmetry is spontaneously broken in
favor of state $a$. The lower curve is $U(ab)$ and $U(bc)$, also
which coincide in the disordered high-temperature phase and split in
the symmetry-broken low-temperature phase. It is seen that the
interface density $U(ab)$, between the symmetry-breaking and
non-symmetry-breaking states, exhibits reentrance as temperature is
lowered in the ordered phase, first increasing in value and then
receding to zero at zero temperature. In Fig. 5, for comparison, the
densities are plotted together for the different $q$ values (and
similarly for the clock models in the adjoining Fig. 6). The
interface density reentrance is pronounced in the low $q$ states,
but continues for high $q$. As temperature is lowered through the phase transition, the reentrance relies on: (1) Due to the increase in the sizes of the domains of the favored stated $s_i=a$, the numbers increase for the $(s_i=a, s_j \neq a)$ pairs at the boundaries of these  domains.  (2) As the sizes of the domains of the favored state $s_i =a$ further increase, upon further lowering the temperature, these domains merge, eliminating the boundaries.  This reentrance is less pronounced for higher $q$, since the phase transition is at lower temperature and (2) sets in before (1) develops.

Reentrance is the reversal of a thermodynamic trend as the system
proceeds along one given thermodynamic direction.  Since its
observation in liquid crystals by Cladis \cite{Cladis}, this
at-first-glance strange phenomenon has attracted attention by the
need for a physical mechanistic explanation, which has been
disparate in disparate systems. Thus, in liquid crystals the
explanation has been the relief of close-packed dipolar frustration
by positional fluctuations (librations) \cite{Netz,Garland}, in
closed-loop binary liquid mixtures the explanation has been the
asymmetric orientational degrees of freedom of the components
\cite{Walker}, in surface adsorption the explanation has been the
buffer effect of the second layer \cite{Caflisch}. In spin-glasses,
where there is orthogonally bidirectional reentrance, the effect of
frustration in both disordering and changing the nature of ordering
(to spin-glass order) is the cause \cite{Ilker1}. In cosmology,
reentrance is due to high-curvature (black hole) gravity
\cite{Mann1, Mann2}. In the current case of Potts (and clock, see
below) interfacial density, in lowering the temperature, when the
system orders in favor state $a$, the preponderance of the latter
also increases its interface with the other states.  However, as
this preponderance further increases and in fact takes over the
system, the other states are eliminated and their interface with $a$
thus is also eliminated.  This happens for all $q$-state Potts and
clock models.

The calculated bond-state densities also readily yield
magnetizations, which will be discussed in Sec. VI, as well as the
different behaviors of the two models in the $q \rightarrow \infty$
limit.

\begin{figure}[ht!]
\centering
\includegraphics[scale=0.23]{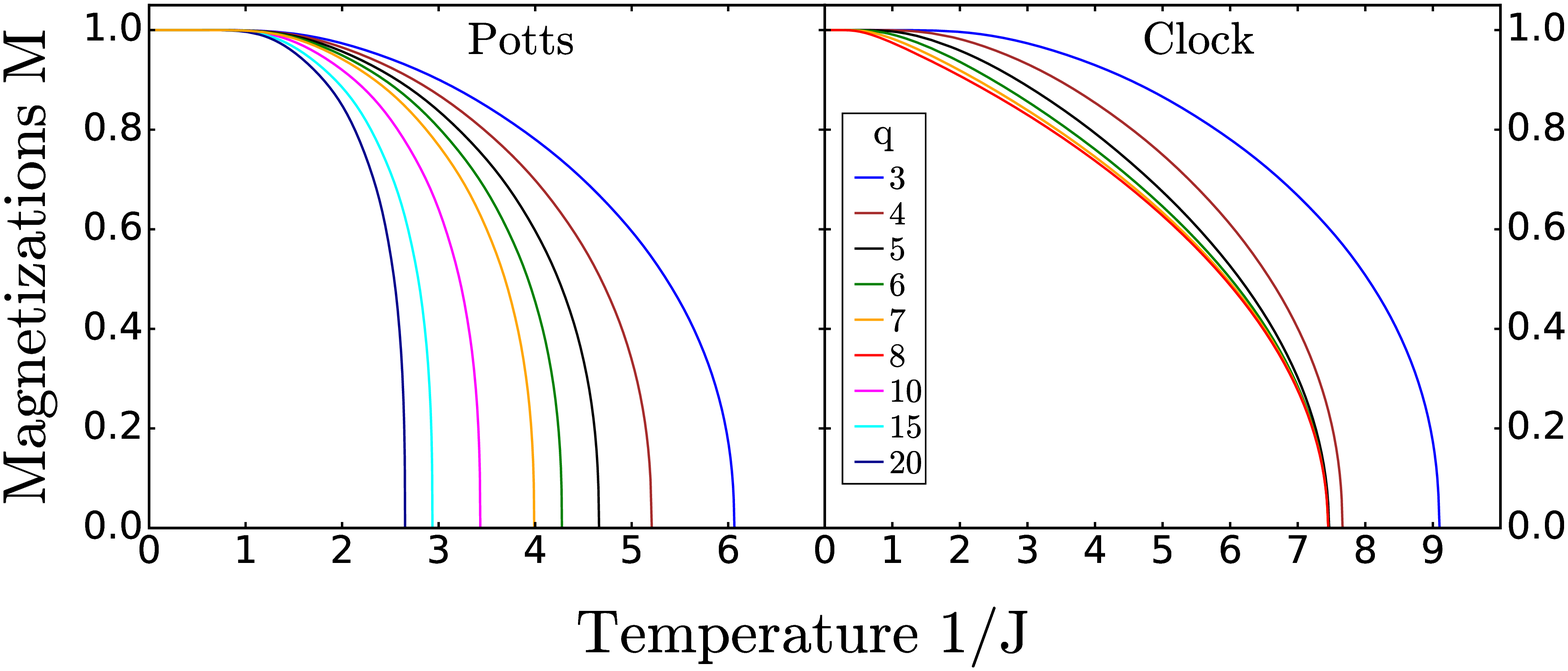}
\caption{Calculated magnetizations of the Potts and clock models as
a function of temperature $J^{-1}$ in $d=3$, for different values of
the number of states $q$. It is noteworthy that, in the clock
models, the magnetization quickly (at as low as $q=5$) settles to
its $q=\infty$ (which is the $XY$ model) value along the entire
temperature range of the low-temperature ordered phase, not only at
the value of $J^{-1}_c$ as was seen above in Fig. 1 and Table I.}
\end{figure}

\section{Results: $q$-State Clock Models}

Clock models do not have permutational symmetry, so that the
recursion relations for the diagonal and the top triangle of the $q
\times q$ energies cannot be reduced to four equations (as in Eqs.
(8) above). Using the different symmetries for each $q$, the number
of these energies that under renormalization group separately recur
can be reduced, but still increases with $q$, eventually numerically
burdening the algebra.

\subsection{Renormalization-Group Calculation and Six-Energy Renormalization-Group Flows for $q=4$}
In the three-state clock model, since with respect to any one state,
the other two states are equivalent, for $q=3$ the clock and Potts
models are identical, up to a factor of $1 - \cos(2\pi/3) = 3/2$ in
the coupling constant $J$.

In the four-state clock model, the six energies that need
be separately recurred under renormalization group are
\begin{multline}
\begin{split}
E(11)= E(33)&,\quad E(22),\quad E(12)=E(23),\\
E(01)= E(03)&,\quad E(02),\quad E(13),\\
\end{split}
\end{multline}
where $E(mn)$ is the energy of neighboring spins with angles $2\pi
m/q$ and $2\pi n/q$. The equalities result from the symmetries of
the $q=4$ state clock model, as the state with $m=0$ is singled out
for possible symmetry breaking.  When, as we do here, the same
energy label is assigned to different states that should have the
same energy by symmetry, the derivative in Eq.~(4) gives the sum of
the densities of these states, as seen below.

The recursion relations are, similarly to Eqs. (8),
\begin{multline}
\begin{split}
e^{E'(mn)+\widetilde{G}}=& \sum_{k=0}^{q-1} \, e^{b^{d-1}E(mk)+b^{d-1}E(kn)},\\
e^{\widetilde{G}} =& \sum_{k=0}^{q-1} \, e^{b^{d-1}E(0k)+b^{d-1}E(k0)}.\\
\end{split}
\end{multline}
The renormalization-group flows and the calculation of the
thermodynamic densities proceed as for the Potts models above.  The
recursion matrix is the $7\times7$ derivative matrix of $[G,E(mn)]$,
where $E(mn)$ are the six energies of Eq.~(15) and $G$ is the
additive constant as in Eq.~(6), a captive variable of the
renormalization-group flows of the $E(mn)$. By derivative matrix, we again mean the derivatives of the renormalized quantities with respect to the unrenormalized quantities.

The left eigenvector with eigenvalue $b^d$ of the recursion matrix
at the phase sinks has the form $[1,\overline{U}(mn)]$, where
$\overline{U}(mn)$ are the density sums conjugate to the recurring
$E(mn)$.  At the high-temperature disordered phase sink, all
energies equal $E(00)$, namely zero, and
$\overline{U}(mn)=z(mn)/q^2$, where $z(mN)$ is the degeneracy of
$E(mn)$, namely $z=2,1,4,4,2,2$ for the energies in Eq.~(15), also
taking into account the degeneracy for label interchange when $m\neq
n$. Repeated application of Eq.~(7) then yields the six density sums
$\overline{U}(mn)$ in the entire temperature range of the disordered
phase. The densities for individual states are obtained from the
sums by $<\delta(mn> = U(mn) = \overline{U}(mn)/z(mn).$  For
example,
\begin{multline}
\begin{split}
\overline{U}(01)/z(01) =& <(\delta(01)+\delta(10)+\delta(03)+\delta(30))>/4 \\
=& <\delta(01)> = U(01).\\
\end{split}
\end{multline}
Thus, when we reduce the number of the recurring energies using
symmetries as in Eq.~(15) and label interchange symmetry, the
renormalization-group calculation yields the density sum
$\overline{U}(01)$, which is then subjected to Eq.~(17).  At the
low-temperature sink, in the left eigenvector with eigenvalue $b^d$,
all $\overline{U}(mn)=0$ and therefore $U(00)= 1- \Sigma_{mn} U(mn)
= 1$, symmetry is broken in favor of state 0. Repeated application
of Eq.~(7) then yields the six density sums $\overline{U}(mn)$ in
the entire temperature range of the ordered phase.  (The other left
eigenvector with eigenvalue $b^d$ is
$\overline{U}(\overline{00})=1$, where $\overline{0}$ is a state
other than 0, and all other recurring $\overline{U}(mn)=0$, giving
an equivalent phase and completing the picture of phase coexistence,
as for the Potts models above.)

The calculated nearest-neighbor densities of the four-state clock
model in $d=3$ are shown in Fig. 3.  The densities
$U(00),U(11),U(33)$ coincide in the disordered high-temperature
phase and, in the low-temperature phase, $U(00)$ splits from
$U(11),U(33)$ under the symmetry breaking favoring the state 0.
Similarly, $U(01),U(03),U(12),U(23)$ coincide in the disordered
high-temperature phase and, in the low-temperature phase,
$U(01),U(03)$ splits from $U(12),U(23)$ under the symmetry breaking.
Similarly, $U(02),U(13)$ coincide in the disordered high-temperature
phase and, in the low-temperature phase, $U(02)$ splits from $U(13)$
under the symmetry breaking.  The densities involving the 0 state
split from their symmetric counterparts in the low-temperature
phase, increasing their values.  This is spontaneous
symmetry breaking. Furthermore, the interface densities
involving the 0 state exhibit reentrance as temperature is lowered
in the ordered phase, first increasing in value and then receding to
zero at zero temperature.

\subsection{Renormalization-Group Flows of Eight, Twelve, Fifteen, Twenty Energies for $q=5,6,7,8$}
The calculations for $q=5,6,7,8$ are more extensive.  Using
symmetries grouping the same values of $|n-m|$, but grouping
separately for positioning with respect to state 0, for the
possibility of spontaneous symmetry breaking, $q=5,6,7,8$ have
renormalization-group flows in eight, twelve, fifteen, twenty
energies, respectively.  These constitute very extensive
renormalization-group calculations.

The results are shown in Fig. 3. Direct comparison between different
$q$ are shown in Fig. 5, showing a striking evolution with respect
to $q$. The characteristic behavior is seen here as well. The curves
are for $U_{k,k-m}$, for $k=0,1,...,q-1$ and $m=0,1,...$. Thus, $m$
measures the angular difference $\theta _i - \theta _j = 2\pi m/q$
between the states of neighboring spins. For each $m$, the curves
for different $k$ coincide in the disordered high-temperature phase.
In the low-temperature phase, for each $m$, the densities involving
$k=0$ and the densities involving $k>0$ split under the spontaneous
symmetry breaking favoring the state 0. The interface densities
involving $k=0$ exhibit reentrance as temperature is lowered in the
ordered phase, first increasing in value and then receding to zero
at zero temperature.

\section{Magnetizations and Infinite $q$ (Non-)Saturation of the Critical Temperature}

The magnetizations $M$ are directly obtained from the
nearest-neighbor densities.  For the Potts models,
\begin{equation}
M = <\delta(s_ia)> = \sum_{m=0,n=0}^{q-1,q-1} U(mn) \, \delta(ma).
\end{equation}
For the clock models,
\begin{equation}
M = <\cos(\theta_i)> = \sum_{m=0,n=0}^{q-1,q-1} U(mn) \, \cos(2\pi
m/q).
\end{equation}
These equations are obtained by including a magnetic field term (to
be taken to zero after differentiating) in the $E(mn)$,
differentiating $\ln(Z)$ with respect to the magnetic field, and
using the chain rule with $E(mn)$ as intermediary.

The results for the magnetizations and the critical temperatures are
given in Figs. 1,7 and Table I. It should be noted that these results are exact for the $d=3$ hierarchical lattice.  They are approximate for the cubic lattice.  Specifically, by allowing effective vacancies to be generated by the renormalization-group transformation, the Potts model transition correctly becomes first order for $q>2$. \cite{BerkerPLG,NienhuisPotts,AndelmanBerker}

It is of interest to see the magnetization curves for the clock models in Fig. 7 settle to their
$q\rightarrow \infty$ value for as low as $q=5$.  This is of course
reflected in the essentially constant value of the clock critical
temperatures as $q$ is increased.

Such is not the case for the Potts models.  Directly writing down
the recursion relation for $J$ in Eq.~(1),
\begin{equation}
 J' = \ln [e^{2 b^{d-1} J}+(q-1)] - \ln [2 e^{b^{d-1} J}+(q-2)],
\end{equation}
setting the fixed point condition $J' = J = J_c$, and expanding for
large $J$ and $q$, we find the critical temperatures
\begin{equation}
J_c^{-1} = 7 / \ln(q).
\end{equation}
This curve is plotted as a dashed curve in Fig. 1 and gives a good
fit even for finite $q$. As $q\rightarrow \infty$ the critical
temperature goes to zero.

There is a physical explanation to the contrast between Potts and
clock.  In the Potts models, the states neighboring $s_i=a$ do not
contribute to the magnetization and are entropically favored as $q$
is increased. (In fact, in the permutationally symmetric Potts
models, the concept of ''neighboring'' state has no meaning: every
state is equally positioned with respect to a chosen state.) By
contrast, in the clock models, the states neighboring $\theta_i=0$
give almost a full contribution, namely $<\cos(2\pi n_i/q)>$ to the
magnetization where $n_i<<q$ for large $q$. Thus, in all spatial
dimensions $d$, the critical temperature for Potts models should go
to zero inverse logarithmically as $q\rightarrow \infty$.

\begin{table}
\begin{tabular}{c c c c c c c c c c c c c c c c}
\hline
q &\vline &Potts $J_c^{-1}$ &\vline &clock  $J_c^{-1}$  &\vline \\
\hline
3  &\vline & 6.062 &\vline &  9.093  &\vline \\
\hline
4 &\vline & 5.206 &\vline &  7.661  &\vline \\
\hline
5 &\vline & 4.660 &\vline &  7.416  &\vline \\
\hline
6 &\vline & 4.277 &\vline &  7.395  &\vline \\
\hline
7  &\vline & 3.990 &\vline &  7.391  &\vline \\
\hline
8 &\vline & 3.764 &\vline &  7.388  &\vline \\
\hline
10 &\vline &3.431  &\vline & 7.384  &\vline \\
\hline
15 &\vline & 2.936 &\vline & 7.381  &\vline \\
\hline
20 &\vline & 2.652 &\vline & 7.379  &\vline \\
\hline
360 &\vline & 1.278 &\vline &  7.377  &\vline \\
\hline
$\infty$ &\vline & $7/\ln(q)$ &\vline &    &\vline \\

\hline \hline
\end{tabular}
\caption{Calculated critical temperatures of $q$-state Potts and
clock models. The clock-model results have been found numerically by doing the sums in the renormalization-group decimation over a very large number of states, up to $q=360$.\\} \label{tab:1}
\end{table}

\section{Conclusion}

In this study, we have calculated all of the bond-state densities of
the $q$-state Potts and clock models in $d=3$.  This was done for
all $q$ for the Potts models, by reducing the recursion relations to
four, using symmetries, modulo singling out one state for possible
spontaneous symmetry breaking, which happens for both models.
Although the number of recursion relations in clock models can be
reduced by symmetry, their number grows, for example to twenty
different energies for our treated eight-state clock model.
However, we have presented a robust method which would make the
calculation for any number of states $q$ feasible.

A reentrant behavior of all of the symmetry-broken interface
densities was found for both models, and physically explained. A
surprising saturation with increasing $q$ was found in the clock
models, but not in the Potts models.  We also found qualitatively
different phase transition behaviors in the $q\rightarrow \infty$
limit, which was physically explained by entropy and alignment
arguments.

\begin{acknowledgments}
Support by the Kadir Has University Doctoral Studies Scholarship
Fund and by the Academy of Sciences of Turkey (T\"UBA) is gratefully
acknowledged.
\end{acknowledgments}

\end{document}